\documentclass[prd,twocolumn,showpacs,preprintnumbers,nofootinbib,amsmath,amssymb,floatfix,aps]{revtex4-1}
\usepackage{url}
\usepackage{graphicx}
\usepackage{subfigure}%
\usepackage{isotope}
\usepackage{color}

\graphicspath{{figs/}}

\begin{document}

\title{Scale Dependence of Nucleon-Nucleon Potentials}
\author{Omar Benhar}
\email{omar.benhar@roma1.infn.it}
\affiliation{INFN and Dipartimento di Fisica, ``Sapienza'' Universit\`a di Roma, I-00185 Roma, Italy}
\date{\today}%

\begin{abstract}
The scale-dependence of the nucleon-nucleon interaction, which in recent years has been extensively analysed within the context of chiral
effective field theory, is, in fact, inherent in any potential models constrained by a fit to scattering data.
A comparison between a purely phenomenological potential and local interactions derived from chiral effective field theory suggests that\textemdash thanks to the ability to describe nucleon-nucleon scattering at higher energies, as well as the deuteron momentum distribution 
extracted from electro-disintegration data\textemdash
phenomenological potentials are best suited for the description of nuclear dynamics at the scale relevant to neutron star matter.
\end{abstract}
{\pacs{13.75.Cs, 21.30.-x, 21.65.-f}}
%

\maketitle


The interpretation of the observed properties of nuclear systems
in terms of forces acting between the constituent nucleons has long been recognised as a formidable endeavor. 
Over six decades after publication of the article in which Hans Bethe famously addressed the question "What Holds the Nucleus Together?"~\cite{Bethe53},  a 
large amount of work is still being devoted to the development of accurate theoretical models of nuclear forces. 
Ideally, such models should provide the basis for a unified description of a broad range of systems, from the deuteron to neutron stars~\cite{bob_rmp}, 
in which nucleon-nucleon interactions occur at different ranges, and therefore involve different energies.

In principle, the description of nuclear dynamics should be based on 
the fundamental theory of strong interactions: Quantum Chromo-Dynamics, or QCD. The efforts aimed at deriving the nucleon-nucleon (NN) potential from lattice QCD have recently 
achieved remarkable success in predicting its prominent qualitative features~\cite{lattice1,lattice2}. 
However, the results of pioneering calculations, performed using potentials obtained from 
lattice QCD studies, suggest that significant developments will be necessary 
to explain nuclear matter properties at quantitative level~\cite{lattice3}.

In the absence of a truly fundamental approach, a number of NN potential models have been developed combining the time-honored Yukawa's 
pion-exchange theory~\cite{Yukawa}\textemdash which is known to describe the interaction at long distance\textemdash and phenomenology. 
In this context, a great deal of empirical information is provided by the two-nucleon system, in both bound and scattering states.
Deuteron properties and the large database of NN scattering phase shifts at laboratory energies up to pion production 
 threshold have been extensively exploited to obtain high-precision phenomenological potentials, see Ref.~\cite{AV18} and references therein.

 A more fundamental approach, in which the nuclear potential is derived from an effective Lagrangian involving pions and 
low-momentum nucleons, constrained by the broken chiral symmetry of QCD, was proposed by Steven Weinberg at the beginning of the 
1990s ~\cite{weinberg1}. 
This formalism provides a systematic scheme, referred to as Chiral Effective Field Theory ($\chi$EFT),  in which nuclear interactions
are expanded in powers of a small parameter, e.g. the ratio between the pion mass, $m_\pi$,  or a typical nucleon momentum, $Q$, and the scale of chiral symmetry breaking, $\Lambda_\chi~\sim~1$~GeV. 
Within this framework, long- and intermediate-range nuclear forces, originating from pion exchange processes, are fully determined by pion-nucleon observables, whereas short-range interactions are described by contact terms involving a set of additional parameters, fixed in such a way as to reproduce NN scattering phase shifts. As pointed out in Ref.~\cite{weinberg1}, a major advantage of $\chi$EFT lies in the 
possibility to describe two- and many-nucleon potentials within a unified formalism.

In this paper, I will focus on the two-nucleon sector, and compare a phenomenological potential to local    
$\chi$EFT interactions, to determine their respective ability to describe nuclear matter at supra-nuclear densities and, more generally, short-range nuclear dynamics. This feature is essential to
neutron-star modelling, and    
will be of paramount importance in the dawning era of gravitational wave astronomy~\cite{GW1,GW2}.

Purely phenomenological potentials such as the Argonne $v_{18}$ model (AV18)~\cite{AV18}, 
widely employed to perform nuclear matter calculations~\cite{apr1,apr2}, are defined in coordinate space in the form
\begin{align}
\label{anlv18}
v_{ij}=\sum_{p=1}^{18} v^{p}(r_{ij}) O^{p}_{ij} \ .
\end{align}
The functions $v^p$, involving a set of adjustable parameters whose value is determined fitting NN data, only depend on the distance between the interacting particles, $r_{ij} = |{\bf r}_i - {\bf r}_j|$.
The operators $O^{p}_{ij}$, on the other hand,  account for the strong spin-isospin dependence of NN interactions, as well as for the presence of non-central forces.
The dominant contributions to the sum appearing in the right-hand side of Eq.\eqref{anlv18} are those associated with the six operators
\begin{align}
O^{p \leq 6}_{ij} = [\openone, (\boldsymbol{\sigma}_{i}\cdot\boldsymbol{\sigma}_{j}), S_{ij}]
\otimes[\openone,(\boldsymbol{\tau}_{i}\cdot\boldsymbol{\tau}_{j})]  \ ,
\label{av18:2}
\end{align}
where the Pauli matrices $\boldsymbol{\sigma}_i$ and $\boldsymbol{\tau}_i$ describe spin and isospin of the $i$-th nucleon, respectively,  and 
 $S_{ij}  =  3 ( {\boldsymbol r}_{ij} \cdot {\boldsymbol \sigma}_i ) ( {\boldsymbol r}_{ij} \cdot {\boldsymbol \sigma}_i )/{\bf r}_{ij}^2 -  (\boldsymbol{\sigma}_{i}\cdot\boldsymbol{\sigma}_{j})$. 
The contributions corresponding to $p~=~7,\ldots,14$ are associated with the non-static components of the NN interaction, while those corresponding to $p=15,\ldots,18$ take into account small violations of charge independence.
For large distances  the AV18 model reduces to Yukawa's one-pion-exchange potential, that can be written in terms of the six operators of Eq.\eqref{av18:2}.  \\

At leading order (LO) of $\chi$EFT, the NN potential comprises Yukawa's one-pion exchange and two contact terms.
Next-to-leading order (NLO) and  next-to-next-to-leading order (N$^2$LO) contributions also involve two-pion exchange, as well as a set of additional contact terms. 

Early $\chi$EFT potentials have  been
derived in momentum space~\cite{EFT1,EFT2}. A procedure to obtain a local coordinate space representation\textemdash needed to carry out  accurate Quantum Monte Carlo (QMC) calculations\textemdash has been developed 
in Refs.~\cite{N2LO,N2LO_PRL}. 
The numerical results of calculations performed using the Auxiliary Field Diffusion Monte Carlo (AFDMC) technique demonstrate that N$^2$LO 
coordinate space Hamiltonians including both two- and three-nucleon interactions provide a remarkably good account of the 
ground-state energies and charge radii of nuclei with $A\leq16$~\cite{chiral_AFDMC}.

Theoretical studies based on $\chi$EFT have been also  extended to nuclear matter~\cite{N2LO,Tews}. However, the present development of the QMC approach, recently reviewed in Ref.~\cite{QMC}, only allows 
to treat pure neutron matter (PNM). Combined analyses of PNM and isospin-symmetric nuclear matter (SNM)\textemdash needed to study the properties of $\beta$-stable neutron star matter\textemdash  have been carried out within the framework of more approximated methods~\cite{Ignazio}. 

Being based on a low momentum expansion, $\chi$EFT is inherently limited when it comes to describing 
dense systems, in which short-range dynamics plays a dominant role and NN interactions involve high energies. In addition, chiral potentials depend on a momentum-space cutoff, $\Lambda$, which in the coordinate-space representation is replaced by a parameter, $R_0$, determining the range of the
regulator function that smoothly cuts off  one- and two-pion exchange interactions at short distances. In Refs.~\cite{N2LO,N2LO_PRL}, the same range, $R_0 \sim 1$ fm, corresponding to $\Lambda \sim 500$~MeV, has been also used to smear 
the $\delta$-functions arising from Fourier transformation of the contact terms.
As a result, $\chi$EFT potentials are expected to describe interactions up to an energy scale\textemdash or, equivalently, down to a resolution scale\textemdash determined by the combined effects of the truncation of the 
low-momentum expansion and the range of the regulator function.

The scale dependence, which naturally emerges within the context of $\chi$EFT, is also
inherent in any phenomenological models of the  NN potential obtained from a fit to scattering data. Because the fit actually involves the  scattering amplitude, which is explicitly energy dependent, in this case the scale is simply determined by the upper limit of the energy range in which the data can be accurately reproduced. Note that this conclusion applies to purely phenomenological and $\chi$EFT potentials alike.  

In view of the above considerations,  the questions arise of what the energy scale relevant to neutron star matter is, 
and what potential model is best suited to describe the corresponding regime.
To address these issues, consider that in strongly degenerate fermion systems, such as cold nuclear matter, only nucleons in states close to the Fermi surface can participate in scattering processes.  
It follows that the center-of-mass energy of the collisions can be written in terms of the nucleon Fermi momentum, which in turn is simply related to the density.
In the case of head-on collisions in PNM at density $n$ one finds
\begin{align}
E_{\rm cm} = \frac{1}{m}  ( 3 \pi^2 n )^{2/3} \  , 
\label{density}
\end{align} 
where  $m$ is the nucleon mass.

\vspace{-.2in}
\begin{figure}[h]
\includegraphics[width=8.5cm]{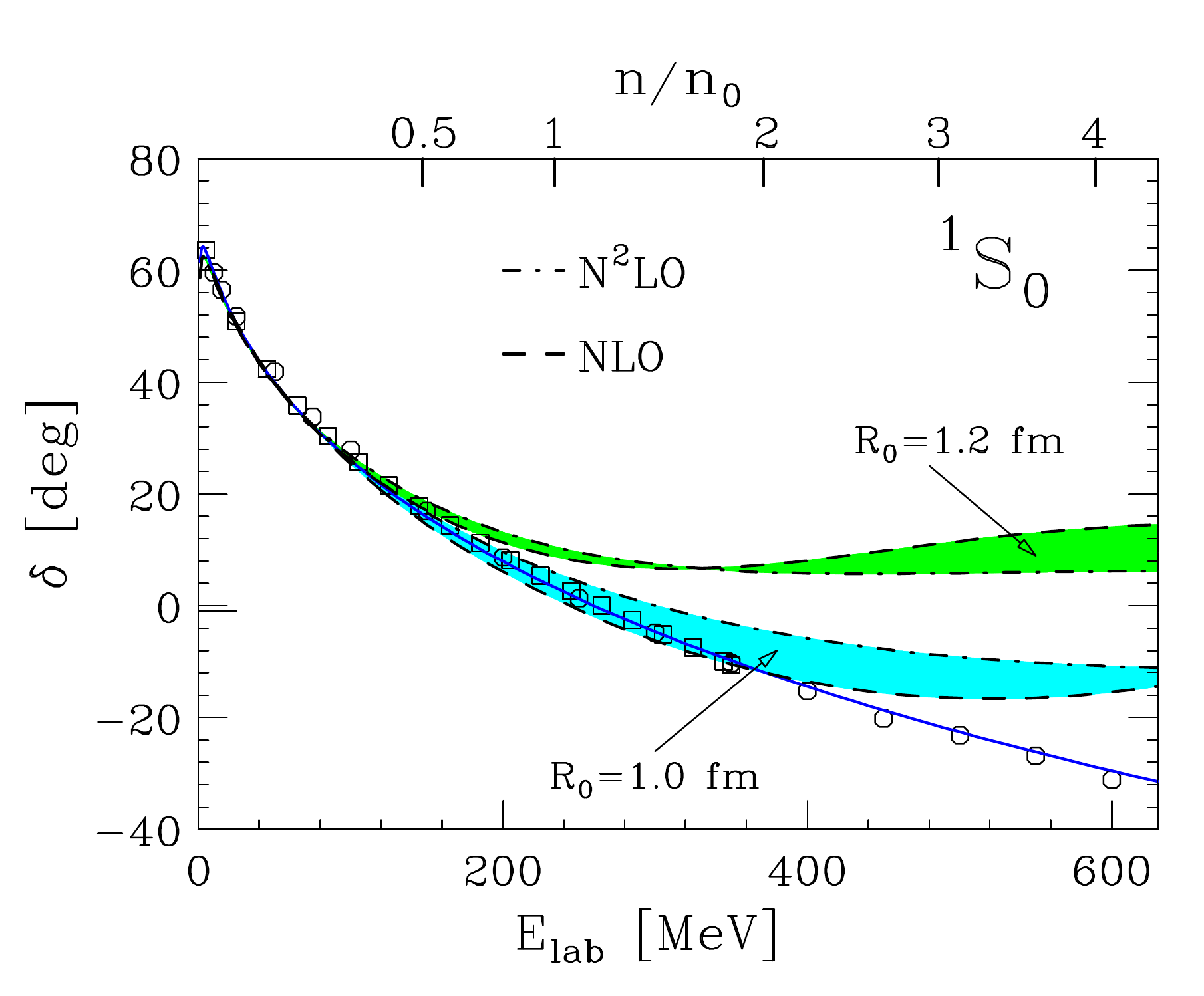}
\vspace{-.1in}
\caption{Neutron-proton scattering phase shitfs in the $^1{\rm S}_0$ channel, as a function of kinetic energy of the beam particle in the
laboratory frame (bottom axis). The corresponding density of PNM\textemdash in units of the equilibrium density
of SNM, $n_0=0.16 \ {\rm fm}^{-3}$\textemdash is given in the top axis. The solid line has been obtained using the AV18 potential, while
the shaded regions illustrate the spread between the NLO (dashed lines) and N$^2$LO (dot-dash lines) predictions of the $\chi$EFT
potentials of Ref.~\cite{N2LO}, obtained setting  $R_0 =$ 1.0 and 1.2 fm. 
Squares and circles represent the results of the analyses of Refs.~\cite{Nijmegen1,Nijmegen2} and~\cite{SAID}, respectively.\label{fig:phases}}
\end{figure}

Figure~\ref{fig:phases} shows the energy dependence of the proton-neutron scattering phase shifts in the $^1$S$_0$ partial wave,  
obtained within the approach of Ref.~\cite{calogero} using the AV18 potential and the local chiral potentials of Ref.~\cite{N2LO}. Note that the AV18 interaction has been 
obtained fitting all phase shifts included in the Nijmegen data base, extending up to $E_{\rm lab} = 350$~MeV, while the fit performed by the 
authors of Ref.~\cite{N2LO} is limited to $E_{\rm lab} =$ 150~MeV and 100~MeV for $R_0 =$ 1.0 and 1.2 fm, respectively.

The solid line represents the results obtained using the AV18 potential, 
while the shaded regions illustrate the spread between the NLO (dashed lines) and N$^2$LO (dot-dash lines) $\chi$EFT predictions corresponding  to $R_0 =$ 1.0 and 1.2 fm. It is apparent that the uncertainty associated with the cutoff  $R_0$ is larger than the one arising form the convergence of the chiral expansion, and that the results obtained with the choice $R_0 = 1$~fm, leading to a harder interaction, provide a better description of the phase shifts at energies larger than $\sim$150~MeV. However, in this case the dashed line, showing the results of the NLO approximation, turns out to lie consistently closer to the data than the dot-dash line, corresponding to the N$^2$LO approximation. 

The top axis of Fig.\ref{fig:phases} reports the values of PNM density, obtained from Eq.~\eqref{density} with $ E_{\rm lab} = 2 E_{\rm cm}$, 
 in units of the equilibrium density 
of SNM, $n_0 = 0.16 \ {\rm fm}^{-3}$. The AV18 potential, yielding an accurate description of the data up to energies 
$E_{\rm lab} \approx 600$~MeV\textemdash well beyond pion production threshold\textemdash appears to be best suited to describe matter at densities as high as $4n_0$, 
while the region of applicability of the  potentials of Ref.~\cite{N2LO} is much more limited. The picture emerging from Fig.~\ref{fig:phases} is consistent with the results of Ref.~\cite{Tews}, whose authors conclude that using interactions obtained from
$\chi$EFT the equation of state of PNM can only be reliably calculated up to $n \lesssim 2n_0$.

The accuracy to which a potential describes short-range interactions, involving large energies, can be further investigated in the two-nucleon sector, 
studying observables that carry information on the high-momentum components of the deuteron wave function.


Up to corrections arising from final state interactions and two-body current contributions, which can be accurately taken into account, 
the measured cross section of the deuteron electro-disintegration process 
\begin{align}
e + \isotope[2][]{H} \to e^\prime + p + n \ , 
\label{react:eep}
\end{align}
in which the scattered electron and the knocked-out proton are detected in coincidence, provides a measurement of the momentum 
distribution 
\begin{align}
\label{nk}
n(k) = \left| \int d^3r e^{i {\bf k} \cdot {\bf r}} \psi_D({\bf r}) \right|^2 \ , 
\end{align}
where $\psi_D({\bf r})$ denotes the deuteron wave function, for $k~\lesssim~300$~MeV~\cite{eep:data,eep}.

Additional information on $n(k)$ is obtained from the cross section of the inclusive reaction, in which only the scattered 
electron is detected~\cite{BDS_RMP}.
The observation of scaling in the variable $y$, defined by the relation
\begin{align}
\label{def:y}
\omega + M_D = \sqrt{ m^2 + (q+y)^2 } +  \sqrt{ m^2 + y^2 } \ , 
\end{align}
where $M_D$ is the deuteron mass and $q$ and $\omega$ denote the momentum and energy transfer, respectively, reflects the onset of the 
kinematical regime in which quasi-elastic nucleon knockout is the dominant process contributing to the cross section.
In this region the electromagnetic response of the target nucleus, which in general depends on both $q$ and $\omega$, becomes
a function of  the single variable $y=y(q,\omega)$, which is simply related to the initial momentum of the struck nucleon, and $n(k)$ can be obtained 
from inclusive data. The analysis carried out by the authors of Ref.~\cite{yscaling}, based on the cross sections reported in Refs.~\cite{ee:data1,ee:data2}, provides an accurate determination of $n(k)$ for momenta as high as 700~MeV. High-momentum components
have been shown to strongly affect the inclusive electron-deuteron cross section in the kinematical region of high $q$ and $\omega \ll \sqrt{q^2 + m^2} - m$, corresponding to large negative values of $y$~\cite{ee:deuteron}. 

\begin{figure}[h]
\includegraphics[width=8.0 cm]{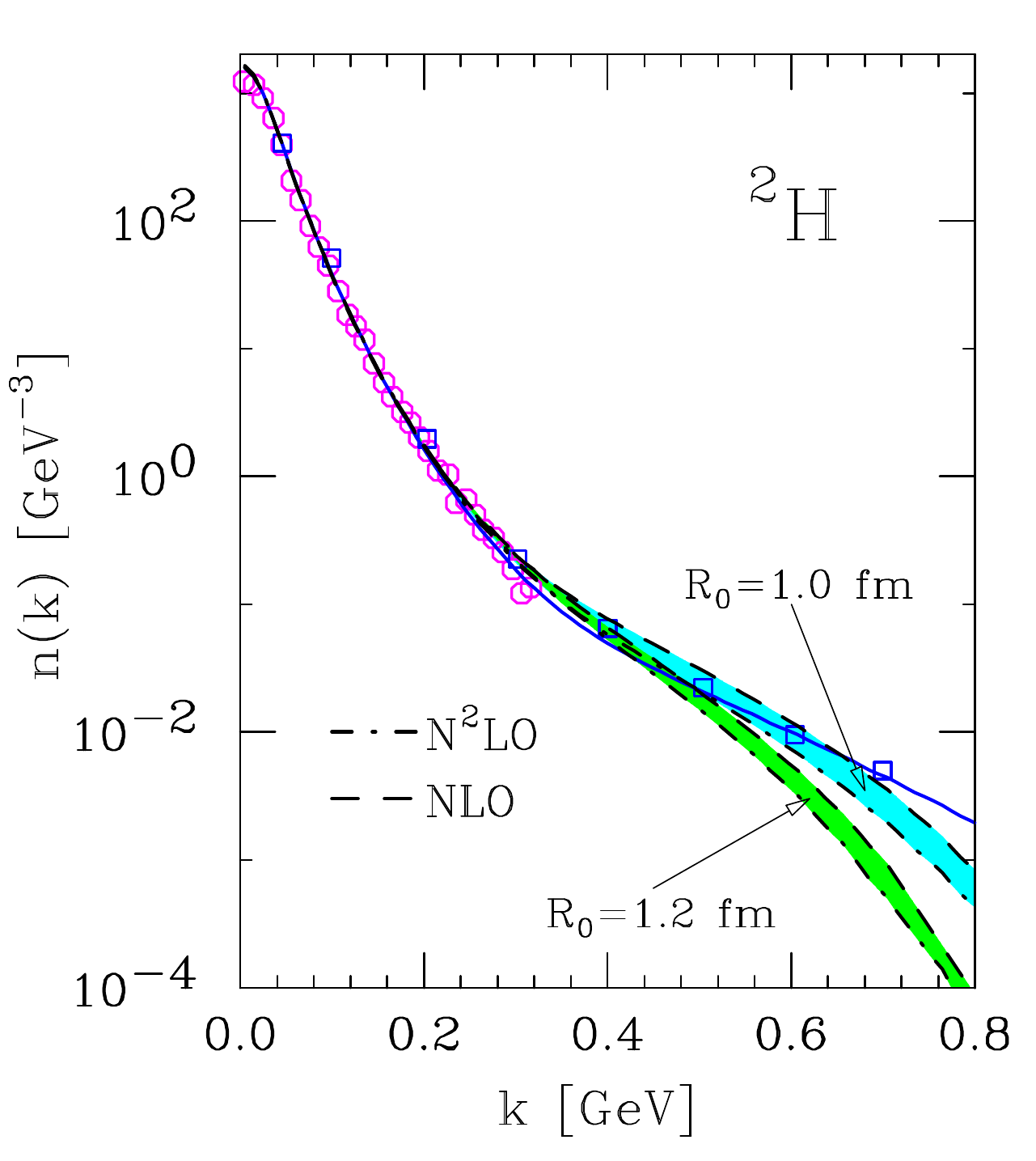}
\vspace{-.1in}
\caption{Deuteron momentum distribution. The solid line has been obtained using the AV18 potential, while
the shaded regions illustrate the spread between the NLO (dashed lines) and N$^2$LO (dot-dash lines) predictions of the $\chi$EFT potentials
of Ref.~\cite{N2LO}, obtained setting  the parameter $R_0$ to 1.0 and 1.2 fm. Open circles and squares represent the results of the analyses of electron scattering data carried out by the authors of Refs.~\cite{eep} and \cite{yscaling}, respectively. \label{fig:momdis}}
\end{figure}

In Fig.~\ref{fig:momdis} the deuteron momentum distributions obtained from the AV18 potential and the chiral potentials of Ref.~\cite{N2LO} are 
compared to the available data. Open circles and squares correspond to the analyses of the measured $\isotope[2][]{H}(e,e^\prime p)$~\cite{eep:data} and $\isotope[2][]{H}(e,e^\prime)$~\cite{ee:data1,ee:data2}
cross sections carried out by the authors of Refs.~\cite{eep} and~\cite{yscaling}, respectively. It has to be emphasized that the excellent agreement between the results of Refs.~\cite{eep} and~\cite{yscaling} at $k \lesssim 300$~MeV strongly supports the validity and  accuracy of the $y$-scaling analysis.
\vspace{-.1in}

The pattern  emerging  from Fig.~\ref{fig:momdis} appears to be consistent with that of Fig.~\ref{fig:phases}. 
The phenomenological AV18 potential accounts for the data up to the highest momentum, while 
the chiral interactions provide an accurate description only at $k \lesssim 300$ MeV. At higher momentum,  
the results obtained using the harder potentials, corresponding to $R_0=1$ fm,  are generally closer to 
the data, but the uncertainty associated with the truncation of the low-momentum expansion 
rapidly increases, reaching $\sim 50$\% at  $k  \sim 500$ MeV. At  $k  \gtrsim 600$ MeV the predictions 
of $\chi$EFT tend to fall well below the experimental points, irrespective of the value of $R_0$, with the disturbing feature that going from NLO to N$^2$LO leads to a sizable
reduction of $n(k)$. To make a connection between the results of Fig.~~\ref{fig:momdis} and nuclear matter properties, consider that 
a Fermi momentum of $\sim$~300 MeV corresponds to a density $n \sim 0.75$  n$_0$  in PNM.

Note that, while accounting for only few percent of the wave-function normalisation, high-momentum components sizeably affect the expectation value of the kinetic energy, $\langle T \rangle$. For example, in the case of the AV18 potential, the region of $k~>~350$~MeV provides about 16\% of the total. Because NN potentials are optimised to reproduce the deuteron ground state energy, $\langle E \rangle$, the availability 
of a momentum distribution providing an accurate description of the data allows for an independent determination of $\langle T \rangle$, which can be exploited
to pin down the expectation value of the interaction energy, $\langle V \rangle$. The values of $\langle E \rangle$, $\langle T \rangle$, and 
$\langle V \rangle$ obtained using the AV18 potential and the local $\chi$EFT potentials of Ref.~\cite{N2LO} are listed in Table~\ref{energies}.
It is apparent that significant differences occur also at the level of average ground state properties.

\begin{table}[h!]
\label{table}
\begin{center}
\begin{tabular}{  l   l  c  c  c }
\hline
\hline
         &           & $\langle E \rangle$  & $\langle T \rangle$ & $\langle V \rangle$ \\
\hline
\multicolumn{2}{ l }{AV18}& -2.225 & 19.791 & -22.016 \\
\hline
    $R_0 = 1.0$ fm  &  LO                 &     \ \     -2.019      \ \                  &     \ \     17.430          \ \              &     \ \        -19.449      \ \                \\ 
                               &  NLO              &     \ \     -2.150       \ \                  &     \ \     20.867          \ \              &     \ \        -23.017      \ \                \\ 
                               &  N$^2$LO      &      \ \     -2.203       \ \                  &     \ \     17.640          \ \              &     \ \        -19.843      \ \                \\ 
\hline
    $R_0 = 1.2$ fm  &  LO                &      \ \     -2.025       \ \                  &     \ \     15.434          \ \              &     \ \          -17.459    \ \                \\ 
                               &  NLO             &      \ \     -2.162        \ \                  &     \ \     16.980          \ \              &     \ \        -19.142      \ \                \\ 
                               &  N$^2$LO      &      \ \     -2.200        \ \                  &     \ \     15.546          \ \              &     \ \        -17.746      \ \                \\                                
\hline
\hline
\end{tabular}
\caption{Breakdown of the deuteron binding energy, $\langle E \rangle$, into kinetic and interaction contributions, denoted 
$\langle T \rangle$ and $\langle V \rangle$, respectively. All energies are given in units of MeV.}
\label{energies}
\end{center}
\end{table}

\vspace{-.2in}

The results discussed in this paper show that, unlike phenomenological models, local potentials derived 
from $\chi$EFT at N$^2$LO level fail to provide an accurate description of NN interactions\textemdash both in free space and in the deuteron\textemdash at a scale corresponding to center-of-mass energies exceeding $\sim 100$ MeV.  As a consequence, phenomenological 
models appear to be best suited to study the properties of dense matter relevant to astrophysical applications\textemdash such as the equation of state 
and the dynamical quantities driving neutron-star evolution and gravitational-wave emission~\cite{NS1,NS2}\textemdash as well as nuclear observables sensitive to short-range  dynamics~\cite{RMP_SRC,SRC2}.

In principle, the ability of $\chi$EFT potentials to describe nuclear interactions at higher energies can be systematically improved taking into account higher order terms of the chiral expansion. However, the results of the state-of-the-art study of Ref.~\cite{EFT3} indicate that including contributions up to N$^4$LO does not dramatically change the picture. The region in which the data is accurately reproduced turns out to be extended to 
$E_{\rm lab} \sim 300$~MeV, corresponding to densities $n \sim1.5 \ n_0$ in PNM, see Fig.\ref{fig:phases}.

Harder $\chi$EFT potentials, suitable to describe dense matter, may be also obtained increasing the 
value of the cutoff $\Lambda$, or, equivalently, reducing the range of the regulator function in coordinate space, $R_0$. However, the authors of Ref.~\cite{N2LO}
report that fitting the NN phase shifts with $R_0 = 0.9$ fm leads to  unnatural values of the couplings associated with contact terms.
In addition, the results of the phase-shift analysis of Ref.~\cite{EFT4}, carried out within the infinite-cutoff renormalisation scheme including terms up to N$^3$LO, show that, while in low angular momentum partial waves cutoff independence is achieved at all orders of the chiral expansion for $\Lambda \gtrsim 5$ GeV,  the expansion does not converge or fails to converge to the experimental data.   

In the absence of a systematic scheme, the uncertainty associated with the phenomenological approach can be estimated comparing 
results obtained from different potentials providing comparable fits of the data, along the line of the work of Ref.~\cite{WFF}.
The results of this study show that the discrepancy between the interaction energies of PNM obtained using the  Argonne $v_{14}$~\cite{AV14} and 
Urbana $v_{14}$~\cite{UV14} NN potentials
is $\sim$ 3\% at $n \sim \ n_0$, and remains $\lesssim$ 20\% up to densities $n \gtrsim 3n_0$.

As a final remark, it must be pointed out that the discussion on the scale dependence of nuclear potentials ultimately brings us to 
the deeper question about the limits of the paradigm underlying nuclear many-body theory. The  description in terms of
point-like nucleons is expected to break down in the neutron star core~\cite{baym}, as well as in scattering processes involving strongly correlated 
nucleons~\cite{annrev,nature}. However, the occurrence of $y$-scaling in electron scattering off a variety of targets, ranging from \isotope[2][]{H} to nuclei 
as heavy as \isotope[197][]{Au}~\cite{yscaling2},  unambiguously shows that at momentum transfer $q \gtrsim 1$~GeV and negative $y$ the beam particles couple to nucleons, carrying momenta up to $\sim 700$~MeV. This observation appears to be supported by the results of a simple model calculation of the properties of the six-quark system~\cite{mark}, suggesting that even in the presence of a significant overlap between nucleons the internal quark structure remains largely unchanged. As a consquence, quantitative studies of the transition to the regime in which degrees of freedom other than nucleons become relevant will require the use of NN interaction models suitable to describe nuclear dynamics up to the scale typical of the  phenomenological approach.


This contents of this paper largely reflect the views expressed in a talk given by the author at the 
Workshop {\em Strong Interaction: From Quarks and Gluons to Nuclei and Stars}, held 
in Erice, Sicily, in September 2018. The kind hospitality of the Ettore Majorana Foundation 
and Centre for Scientific Culture is gratefully acknowledged. Thanks are also due to A. Lovato, 
A. Polls, J. Wambach and R.B. Wiringa for their critical reading of the manuscript.



\begin{thebibliography}{99}

\bibitem{Bethe53}
H.A. Bethe, Scientific American {\bf 189}, 58 (1953).

\bibitem{bob_rmp}
R.B. Wiringa, Rev. Mod. Phys. {\bf 65}, 231 (1993).

\bibitem{lattice1}
N.~Ishii, S.~Aoki, and T.~Hatsuda, Phys. Rev. Lett. {\bf 99}, 022001 (2007).

\bibitem{lattice2}
S. Aoki {\em et al} (HAL QCD Collaboration), PTEP {\bf 2012},  01A105 (2012).

\bibitem{lattice3}
J.~Hu, H.~Toki, and H.~Shen, Scientific Reports {\bf 6}, 35590 (2016).

\bibitem{Yukawa}
H. Yukawa, Proc. Phys. Math. Soc. Jpn. {\bf 17}, 48 (1935).

\bibitem{AV18}
R.B. Wiringa, V.G.J. Stoks, and R. Schiavilla, Phys. Rev. C {\bf 51}, 38 (1995).

\bibitem{weinberg1}
S.~Weinberg,  Phys. Lett. B, {\bf 251}, 288 (1990).

\bibitem{GW1}
B.P. Abbott {\em et al} (LIGO Scientific Collaboration and Virgo Collaboration)
Phys. Rev. Lett. {\bf 119}, 161101 (2017).


\bibitem{GW2}
B.P. Abbott {\em et al} (The LIGO Scientific Collaboration and the Virgo Collaboration)
Phys. Rev. Lett. {\bf 121}, 161101 (2018). 


\bibitem{apr1}
A. Akmal and V.R. Pandharipande, Phys. Rev. C {\bf 56},  2261 (1997) .

\bibitem{apr2}
A. Akmal, V.R. Pandharipande,  and D.G. Ravenhall, Phys. Rev. C {\bf 58},  1804 (1998).


\bibitem{EFT1}
E. Epelbaum, H.W. Hammer, and U.G. Mei{\ss}ner, Rev. Mod. Phys. {\bf 81}, 1773 (2009).

\bibitem{EFT2}
R. Machleidt, and D.R. Entem, Phys. Rep. {\bf 503}, 1  (2011).

\bibitem{N2LO_PRL}
A. Gezerlis {\em et al}, Phys. Rev. Lett. {\bf 111}, 032501 (2013).

\bibitem{N2LO}
A. Gezerlis {\em et al}, Phys. Rev. C {\bf 90}, 054323 (2014).

\bibitem{chiral_AFDMC}
D. Lonardoni {\em et al}, Phys. Rev. Lett. {\bf 120}, 122502 (2018).

\bibitem{Tews}
I. Tews, J. Carlson, S. Gandolfi, and S. Reddy, ApJ {\bf 860}, 149 (2018).

\bibitem{QMC}
J. Carlson {\em et al}, Rev. Mod. Phys. {\bf 87}, 1067 (2015).

\bibitem{Ignazio}
I. Bombaci and D. Logoteta, A\&A {\bf 609}, A128 (2018).




\bibitem{calogero}
F. Calogero and D.G. Ravenhall, Nuovo Cimento {\bf 32} 1755 (1964).

\bibitem{Nijmegen1}
J.R. Bergervoet {\em et al}, 
Phys. Rev. C {\bf 41}, 1435 (1990).

\bibitem{Nijmegen2}
V.G.J. Stoks, R. A. M. Klomp, M.C.M. Rentmeester, and J.J. de Swart, Phys. Rev. C {\bf 48}, 792 (1993).

\bibitem{SAID}
R.L. Workman, W.J. Briscoe, and I.I. Strakovsky, Phys. Rev. C {\bf 94}, 065203 (2016).

\bibitem{eep:data}
M. Bernheim {\em et al}, Nucl. Phys. A {\bf 365}, 349 (1981).

\bibitem{eep}
H. Arenh\"ovel, Nucl. Phys. A {\bf 384}, 287 (1982).

\bibitem{BDS_RMP}
O. Benhar, D. Day and I. Sick, Rev. Mod. Phys. {\bf 80}, 189 (2008).

\bibitem{yscaling}
C. Ciofi degli Atti, E. Pace, and G. Salm\`e, Phys. Rev. C {\bf 36}, 1208 (1987).

\bibitem{ee:data1}
W. Schutz {\em et al}, Phys. Rev. Lett. {\bf 38}, 259 (1977).

\bibitem{ee:data2} 
S. Rock {\em et al}, Phys. Rev. Lett. {\bf 49}, 1139 (1982).

\bibitem{ee:deuteron} 
O. Benhar and V.R. Pandharipande, Phys. Rev. C {\bf 47}, 2218 (1993).

\bibitem{NS1}
O. Benhar and A. Lovato, Phys. Rev. C {\bf 96}, 054301 (2017).

\bibitem{NS2}
G. Camelio, A. Lovato, L. Gualtieri, O. Benhar, J.A. Pons, and V. Ferrari, Phys. Rev. D {\bf 96}, 043015 (2017).

\bibitem{RMP_SRC}
O. Benhar, S.C. Pieper, and V.R. Pandharipande, Rev. Mod. Phys. {\bf 65}, 817  (1993).

\bibitem{SRC2}
N. Fomin, D. Higinbotham, M. Sargsian, and P. Solvignon, Ann. Rev. Nucl. Part. Sci. {\bf 67}, 129 (2017).

\bibitem{EFT3}
E. Epelbaum, H. Krebs, and U.G. Mei{\ss}ner, Phys. Rev. Lett. {\bf 115}, 122301 (2015).

\bibitem{EFT4}
Ch. Zeoli, R. Machleidt, and D.R. Entem, Few-Body Syst. {\bf 54}, 2191 (2013).

\bibitem{WFF}
R.B. Wiringa, A. Fabrocini, and V. Fiks, Phys. Rev. C {\bf 38} 1010 (1988).

\bibitem{AV14}
R.B. Wiringa, R.A. Smith, and T.L. Ainsworth, Phys. Rev. C {\bf 29}, 1207 (1984).

\bibitem{UV14}
I. E. Lagaris and V.R. Pandharipande, Nucl. Phys. A {\bf 359}, 331 (1981).





\bibitem{baym}
G. Baym, T. Hatsuda, T. Kojo, P.D. Powell, Y. Song, and T. Takatsuka, Rep. Prog. Phys. {\bf 81}, 056902 (2018).

\bibitem{annrev}
N. Fomin, D. Higinbotham, M. Sargsian, and P. Solvignon, 
Annu. Rev. Nucl. Part. Sci. {\bf 67}, 129 (2017).

\bibitem{nature}
B. Schmookler {\em et al} (The CLAS Collaboration), Nature {\bf 566}, 354 (2019).

\bibitem{yscaling2}
J. Arrington {\em et al}, Phys. Rev. Lett. {\bf 82}, 2056 (1999).

\bibitem{mark}
M.W. Paris and V.R. Pandharipande, Phys. Rev. C {\bf 62}, 015201 (2000).












\end{thebibliography}
\end{document}